\documentclass[twocolumn,prl,amsart,showpacs,preprintnumbers,amsmath,amssymb]{revtex4-1}

\usepackage[T1]{fontenc}
\usepackage[latin9]{inputenc}
\usepackage[english]{babel}
\usepackage{graphicx}% Include figure files
\usepackage{dcolumn}% Align table columns on decimal point
\usepackage{bm}% bold math
\usepackage{bbm}
\usepackage{hyperref}
\usepackage{color}
\usepackage{mathtools}

\usepackage{amsmath}
\usepackage{mathtools}
\usepackage{cancel}
\usepackage{color}

\begin{document}
\title{Measuring the magnetic moment density in patterned ultrathin ferromagnets\\
 with submicron resolution}
\author{T. Hingant$^{1,\dag}$, J.-P.~Tetienne$^{1,\dag}$, L. J. Mart\'inez$^{1}$, K.~Garcia$^{2}$, D.~Ravelosona$^2$, J.-F. Roch$^{1}$ and V.~Jacques$^{1}$}
\email{vjacques@ens-cachan.fr}
\affiliation{$^{1}$Laboratoire Aim\'e Cotton, CNRS, Universit\'e Paris-Sud and ENS Cachan, 91405 Orsay, France \\
$^{2}$Institut d'Electronique Fondamentale, Universit\'e Paris-Sud and CNRS UMR 8622, 91405 Orsay, France}
\altaffiliation{These authors contributed equally to this work.}

\begin{abstract}   
We present a new approach to infer the surface density of magnetic moments $I_s$ in ultrathin ferromagnetic films with perpendicular anisotropy. It relies on quantitative stray field measurements with an atomic-size magnetometer based on the nitrogen-vacancy center in diamond. The method is applied to microstructures patterned in a 1-nm-thick film of CoFeB. We report measurements of $I_s$ with a few percent uncertainty and a spatial resolution in the range of $(100$~nm)$^2$, an improvement by several orders of magnitude over existing methods. As an example of application, we measure the modifications of $I_s$ induced by local irradiation with He$^+$ ions in an ultrathin ferromagnetic wire. This method offers a new route to study variations of magnetic properties at the nanoscale.
\end{abstract}

%This experiment demonstrates how the method can be used offers a new route to study modifications of magnetic properties at the nanoscale.

%we show how the method can be used to study the effect of local irradiation with He$^+$ ions on the saturation magnetization of an ultrathin ferromagnetic wire
\maketitle

Ultrathin ferromagnetic films with perpendicular magnetic anisotropy (PMA) have attracted considerable interest over the last years both for fundamental studies in nanomagnetism and for the development of a new generation of low power spintronic devices~\cite{Ikeda2010,Parkin2008,Thiaville2012}. In this context, it is crucial to determine accurately the surface density of magnetic moments $I_s$ in such ultrathin ferromagnets, which can not be simply inferred from tabulated bulk values owing to significant interface effects~\cite{Ryu2014}. To this end, macroscopic magnetometry methods like superconducting quantum interference devices (SQUIDs) or vibrating-sample magnetometers have become ubiquitous owing to simplicity of use. However, these conventional techniques are prone to parasitic magnetic signals, leading to an intrinsic background on the order of $10^{-10}$~A.m$^2$~\cite{Ney2008,Garcia2009,Pereira2011}, which corresponds to $\approx 10^{13} \mu_B$~\cite{Note}. To overcome this background, the ferromagnetic signal needs to be averaged over a large sample, thus limiting drastically the spatial resolution of the measurement. As an example, we consider a ferromagnetic film with a typical thickness $t=1$~nm and a saturation magnetization of $M_s=10^6$~A/m, corresponding to a surface density of magnetic moments $I_s=M_s t \approx 100 \ \mu_B /{\rm nm}^2$. In order to reach a signal-to-background ratio of $\sim 10$, the signal must be averaged over a surface larger than  $(1 \ {\rm mm})^2$.\\
\indent Different approaches have been used to tackle this issue. Notably, a recent improvement of the spatial resolution down to the order of $(10 \ \mu {\rm m})^2$ has been achieved by measuring the dipolar repulsion of magnetic domain walls with a magneto-optical Kerr microscope~\cite{Vernier2014}. The use of micro- and even nano-SQUID also allows for a significant gain in sensitivity and resolution. However it remains difficult with such devices to determine accurately the link between magnetic flux and magnetization, and thus to extract precise values of $I_s$~\cite{Wernsdorfer2009}.\\
\indent In this work, we use a single nitrogen-vacancy (NV) center in diamond as an atomic-size magnetometer to probe the stray field above ultrathin ferromagnets and infer $I_s$ on a scale smaller than $(100~{\rm nm})^2$. Importantly, the experiment operates under ambient conditions without applying any external magnetic field, which enables excluding parasitic signals from extrinsic magnetic impurities. As an example of application, we demonstrate how this method can be used to study local variations of magnetic properties with submicron resolution, by measuring modifications of $I_s$ induced by local irradiation with He$^+$ ions in an ultrathin ferromagnetic wire.

\begin{figure}[t]
\begin{center}
\includegraphics[width=0.51\textwidth]{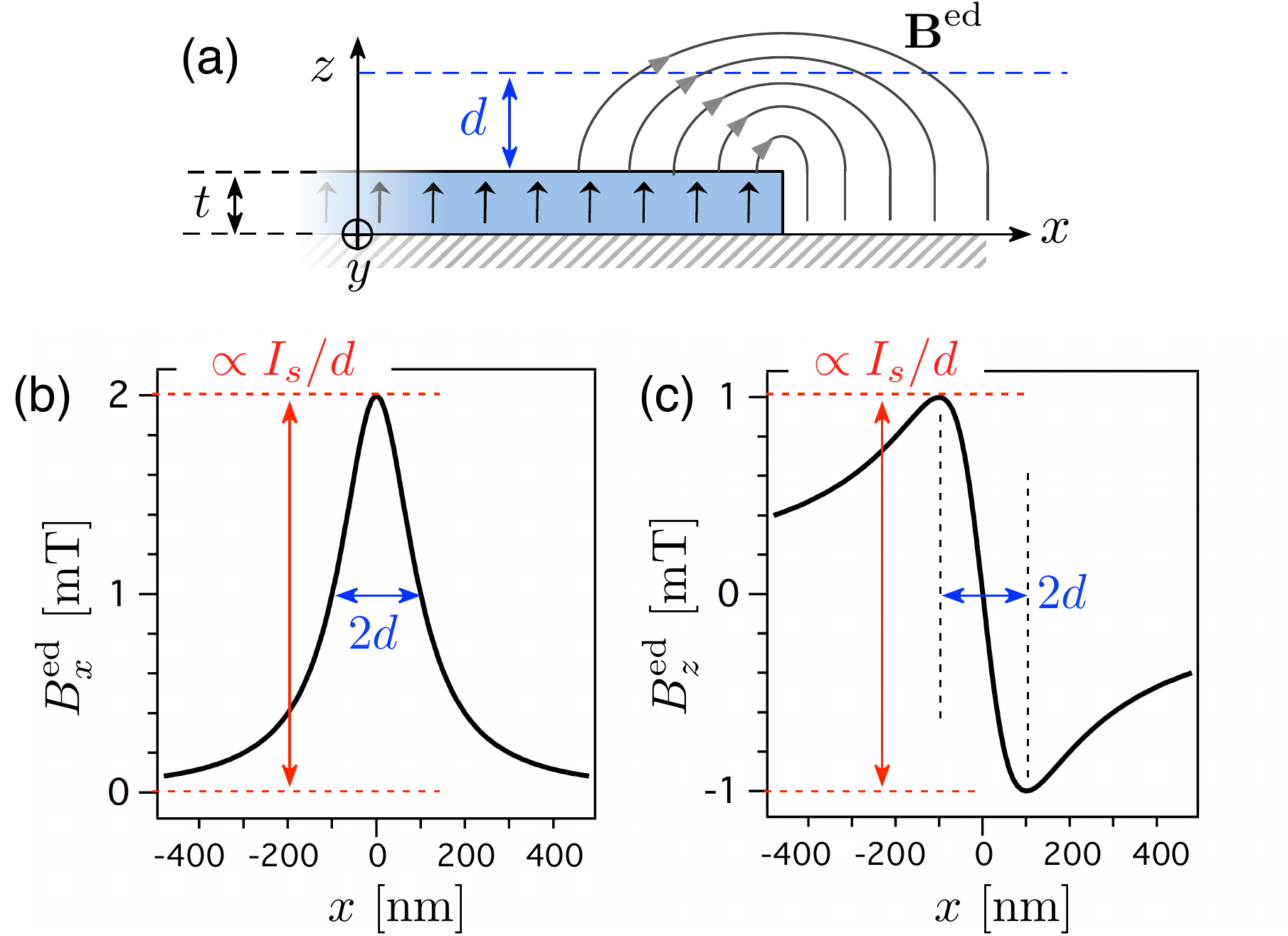}
\caption{(a) Schematic view of the magnetization ${\bf M}$ at the edge of a ferromagnetic material with PMA. The grey arrows indicate the resulting magnetic field lines. (b),(c) Magnetic field components $B^{\rm ed}_x (x)$ and $B^{\rm ed}_z (x)$ calculated at a distance $d=100$~nm using Eq.~(\ref{eq:B_edge}) for $M_s=10^6$~A/m and $t=1$~nm, corresponding to $I_s=M_s t \approx 100 \ \mu_B /{\rm nm}^2$, which is a typical value for the ferromagnetic samples considered in this work. The edge is placed at $x=0$.}
\label{Fig1}
\end{center}
\end{figure}

\begin{figure*}[t]
\begin{center}
\includegraphics[width=1.01\textwidth]{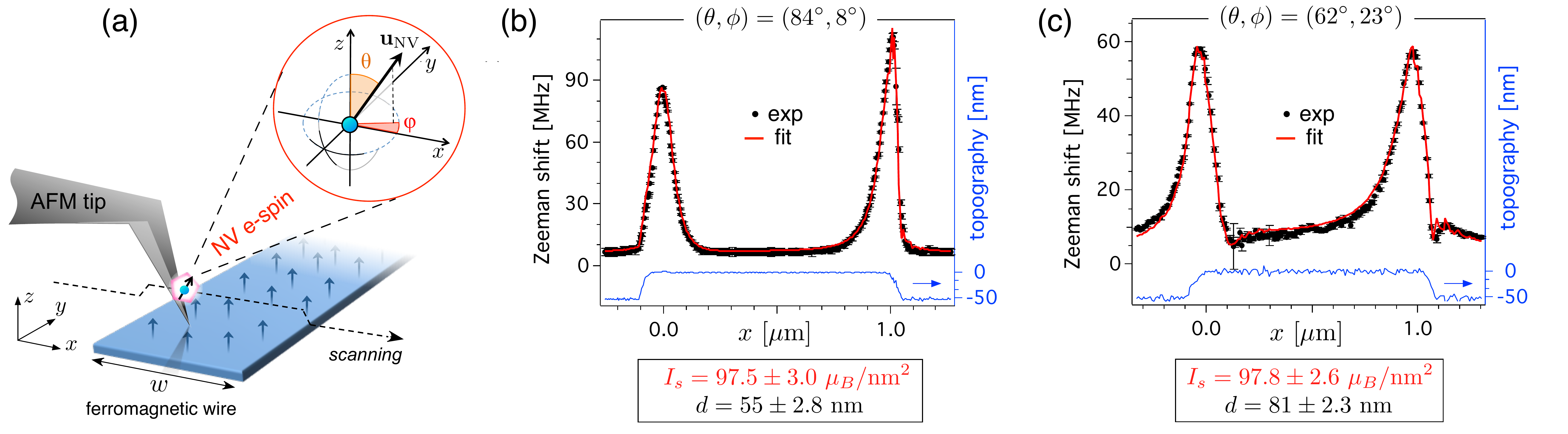}
\caption{(a) Principle of the experiment. Magnetic field measurements across a ferromagnetic wire (width $w$) are performed with a scanning-NV magnetometer operating in tapping mode. The inset indicates the quantization axis of the NV defect electronic spin $\bf{u}_{\rm NV}$, characterized by the spherical angles $\theta$ and $\phi$ in the $(xyz)$ reference frame. (b),(c) Zeeman-shift profile measured by scanning the NV defect across a 1-$\mu$m-wide wire of Ta|CoFeB (1 nm)|MgO. The blue curve is the simultaneously recorded AFM topography of the sample. Two NV probes with different orientations $(\theta,\phi)$ are used in (b) and (c). The red solid lines are data fitting from which the parameters $(I_s,d)$ are extracted. The uncertainties, corresponding to one standard deviation, are evaluated by following the analysis described in the main text.}
\label{Fig2}
\end{center}
\end{figure*}

%Such magnetic charges are the source of a stray magnetic field that scales linearly with the number of surrounding magnetic moments. the source of a radiated magnetic field. 

%on a stray magnetic field that scales linearly with the number of surrounding magnetic moments. 

%In an uniformly magnetized ferromagnetic film, magnetic charges produce a stray magnetic field only at the edge of the film, in a same way that an electric field is only generated at the edge of a planar capacitor. 

We start by describing the general principle of the method. Any magnetization pattern presenting a non-zero divergence ${\bf \nabla}\cdot{\bf M} \neq 0 $ produces magnetic charges with opposite signs, which play a role similar to electric charges in electrostatics. Therefore, a magnetic film with PMA can be seen as the magnetic counterpart of a planar capacitor. In a same way that an electric field is only generated out of the edges of a planar capacitor, magnetic field is only produced at the edge of an uniformly magnetized ferromagnetic film. The central idea of this work is to directly infer the surface density of magnetic moments $I_s$ through local and quantitative measurements of this stray field, denoted ${\bf B}^{\rm ed}$. For an ultrathin magnetic film with PMA, it scales linearly with the number of surrounding magnetic moments and can be computed analytically at any distance $d$ above the edge. In the limit $d\gg t$ and considering a one-dimensional (1D) model with an infinitely long edge along the $y$ axis [Fig.~\ref{Fig1}(a)], the stray field components above an edge placed at $x=0$ are simply given by~\cite{SI}
\begin{equation}
\begin{dcases} 
B^{\rm ed}_x (x)= \ \ \frac{\mu_0I_s}{2\pi}\frac{d}{x^2+d^2} + O((t/d)^3) \\
B^{\rm ed}_z (x)= -\frac{\mu_0I_s}{2\pi}\frac{x}{x^2+d^2} + O((t/d)^3) \ .
\end{dcases}
\label{eq:B_edge} 
\end{equation} 
For both components, the field maximum scales as $I_s/d$ while the characteristic width of the distribution is given by $2d$ [see Figs.~\ref{Fig1}(b) and (c)]. The value of $I_s$ and $d$ can therefore be directly inferred by using Eq.~(\ref{eq:B_edge}) to fit magnetic field distributions recorded above the edge of an uniformly magnetized ferromagnetic film. 

Although the principle of the method is rather simple, its experimental realization is highly demanding since it requires quantitative magnetic field measurements combined with a spatial resolution at the nanoscale. To meet these requirements, we employ a recently introduced magnetometry technique based on a single NV defect in diamond~\cite{Taylor2008,Balasubramanian2008,Rondin2012}. This point-like impurity has a spin triplet ground state whose electron spin resonance (ESR) can be interrogated by optical means~\cite{Gruber_Science1997}. This property enables quantitative magnetic field measurements within an atomic-size detection volume by recording Zeeman shifts of the NV defect electronic spin sublevels~\cite{Rondin2014}. In the last years, NV-based magnetometry has been used to investigate magnetic vortices~\cite{Rondin2013} and spin-wave excitations in ferromagnetic microdiscs~\cite{Yacoby2014}, as well as domain walls in ultrathin ferromagnets~\cite{Tetienne2014} and bio-magnetism~\cite{LeSage2013}. In the present study, a single NV defect hosted in a diamond nanocrystal is grafted at the apex of an atomic force microscope (AFM) tip and scanned across the edge of an uniformly magnetized ferromagnetic film [Fig.~\ref{Fig2}(a)]. At each point of the scan, the stray field is encoded into a Zeeman shift $\Delta f_{NV}$ of the NV defect ESR frequency, which is well approximated by $\Delta f_{NV} =  \sqrt{E^2+(\gamma_e B_{\rm NV}/2\pi)^2}$ in the magnetic field range considered in this work ($<5$~mT)~\cite{Rondin2014}. Here $E$ is the transverse zero-field splitting parameter of the NV sensor which is typically on the order of few MHz, $\gamma_e/2\pi \approx 28$ GHz/T is the electronic spin gyromagnetic ratio and $B_{\rm NV}$ is the magnetic field projection along the NV defect quantization axis ${\bf u}_{\rm NV}$, defined by the spherical angles $(\theta,\phi)$ in the laboratory frame of reference $(x,y,z)$ [Fig.~\ref{Fig2}(a)]. These angles are measured independently by recording the ESR frequency as a function of the amplitude and orientation of a calibrated magnetic field while the $E$ parameter is precisely inferred by recording an ESR spectrum in zero field~\cite{Rondin2013} .

As a first experiment, scanning-NV magnetometry was used to infer the magnetic moment density in a 1-nm-thick film of CoFeB with PMA. More precisely, we studied a multilayer stack of Ta(5 nm)|CoFeB (1~nm)|MgO(2 nm)|Ta(5 nm) deposited by sputtering on a Si/SiO$_{2}$ substrate~\cite{SI,Burrowes2013}. The film was patterned into 1-$\mu$m-wide magnetic wires using e-beam lithography followed by ion beam etching. The stray field across the magnetic wire then reads
\begin{equation}
{\bf B}^{w}(x)={\bf B}^{\rm ed}(x)-{\bf B}^{\rm ed}(x+w) \ ,
\label{EqWire}
\end{equation}
where the field components of ${\bf B}^{\rm ed}(x)$ are given by Eq.~(\ref{eq:B_edge}) and $w$ is the wire width [Fig.~\ref{Fig2}(a)]. Using a wire rather than a single edge provides a more reliable distance reference along the $x$ axis, which enables us to correct any systematic error caused by the calibration of the AFM scanner. A typical Zeeman-shift profile recorded while scanning the NV defect through the edges of the wire is shown in Fig.~\ref{Fig2}(b) together with the simultaneously recorded topography of the sample. Here the Zeeman shift results from the projection of ${\bf B}^{w}$ along the NV axis. Using a NV probe with a different orientation $(\theta,\phi)$ therefore leads to a modified Zeeman-shift profile, as illustrated in Fig.~\ref{Fig2}(c). We note that the dissymmetry between the field maxima above the two edges is linked to the topography of the sample and the precise position of the NV defect with respect to the end of the AFM tip~\cite{SI}.

\begin{figure}[t]
\begin{center}
\includegraphics[width=0.48\textwidth]{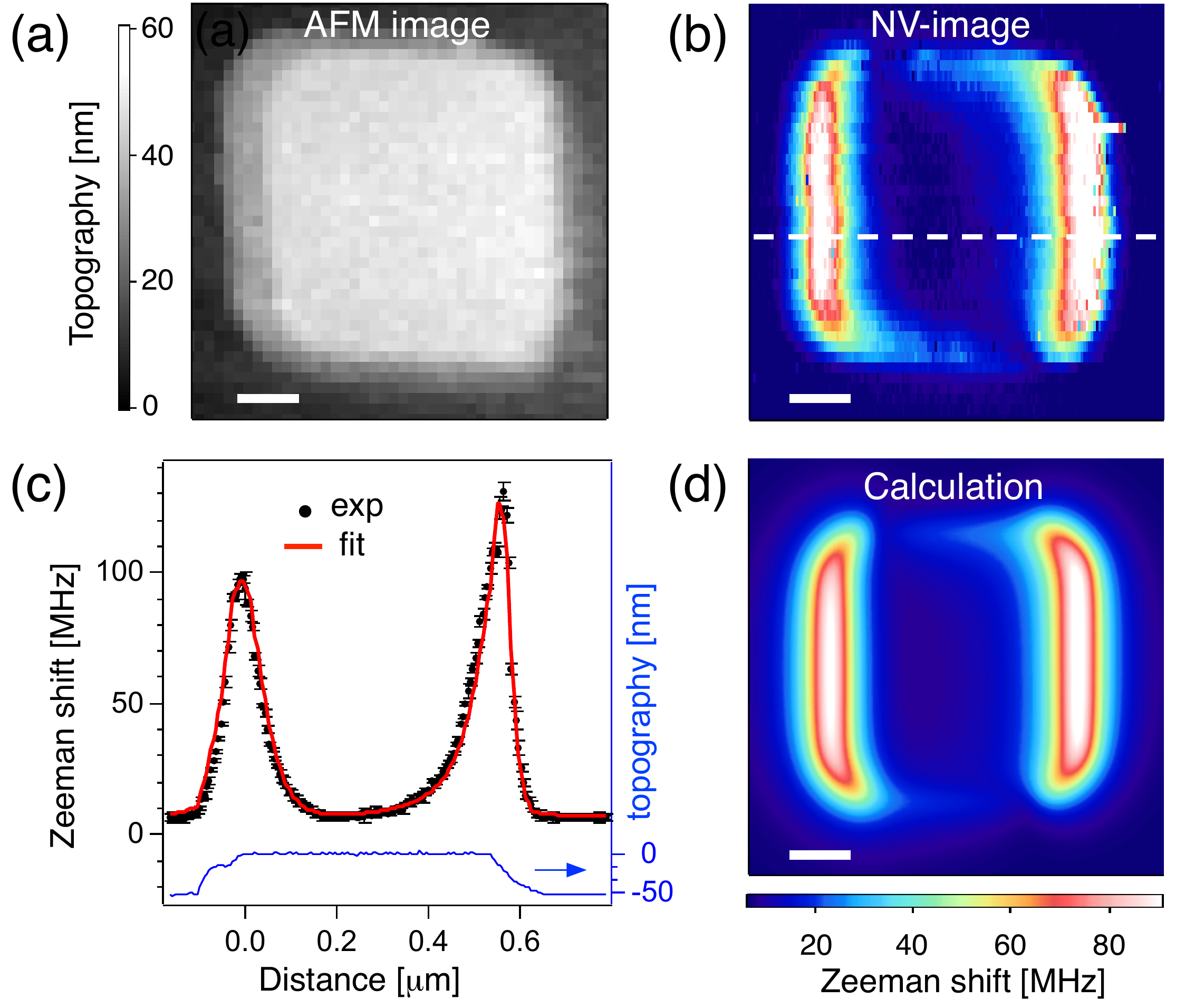}
\caption{(a) AFM image and (b) corresponding Zeeman-shift distribution recorded while scanning an NV center above a square dot etched in a Ta|CoFeB (1 nm)|MgO film. Scale bar:~$100$~nm. (c) Linecut extracted from the white dashed line in (b) together with the simultaneously recorded topography of the sample (blue curve). The red solid line is data fitting, yielding $I_s = 98\pm 3 \ \mu_B /{\rm nm}^2$ and $d=54\pm3$~nm. (d) Calculation of the full Zeeman-shift distribution above the square dot using these parameters. Scale bar: $100$~nm.
}
\label{Fig3}
\end{center}
\end{figure}

The experimental data were then fitted to Eq.~(\ref{EqWire}) with $I_s$ and $d$ as free parameters, while taking into account the topography of the sample~\cite{SI}. The results of the fit are indicated as red solid lines in Figs.~\ref{Fig2}(b) and (c), showing a very good agreement with experimental data. To analyze the precision of the fitting procedure, a statistic of the fit outcomes ($I_s,d$) was obtained for each NV probe by fitting a set of $\approx 20$ independent measurements, leading to a standard deviation smaller than $1\%$. Uncertainties induced by those on the NV defects characteristics $(\theta,\phi,E)$ and the magnetic wire geometry $(w)$ were carefully analyzed by following the method described in detail in Ref.~\cite{Tetienne2014bis}. This leads to an overall uncertainty of the fit outcomes ($I_s,d$) on the order of a few percents. We stress that experiments performed with different NV probes yield to identical results for $I_s$, which further illustrates the robustness of the method [Figs.~\ref{Fig2}(b),(c)]. For a 1-nm-thick film of CoFeB with PMA, we obtain $I_s = 97 \pm 3 \ \mu_B/{\rm nm}^2$. This value is in good agreement with measurements performed on the same films with conventional SQUID magnetometry~\cite{Vernier2014}. 

The main advantage of our approach over existing methods is the large gain in spatial resolution. Indeed, by probing the field in close vicinity of the sample, quantitative values of $I_s$ can be obtained locally as long as the field maxima above the edges can be spatially resolved. As shown in Fig.~\ref{Fig1}, the lateral spread of the stray field is on the order of $d$. The further the probe, the wider the field features, which is clearly visible in Figs.~\ref{Fig2} (b) and (c). The spatial resolution of $I_s$ measurement is therefore on the order of $d$, which is the range of $50$-$100$~nm in this work. This corresponds to an improvement by at least four orders of magnitude over other existing methods~\cite{Vernier2014}. We note that in the present work the probe-to-sample distance $d$ is limited by (i) the size of the diamond nanocrystal ($\sim 50$~nm) and (ii) its imperfect positioning at the apex of the AFM tip~\cite{Tetienne2013}. Further improvement of the spatial resolution down to $d\sim10$~nm could be achieved by employing a single NV defect hosted in all-diamond scanning probe tips~\cite{Maletinsky2012}.

\begin{figure*}[t]
\begin{center}
\includegraphics[width=0.97\textwidth]{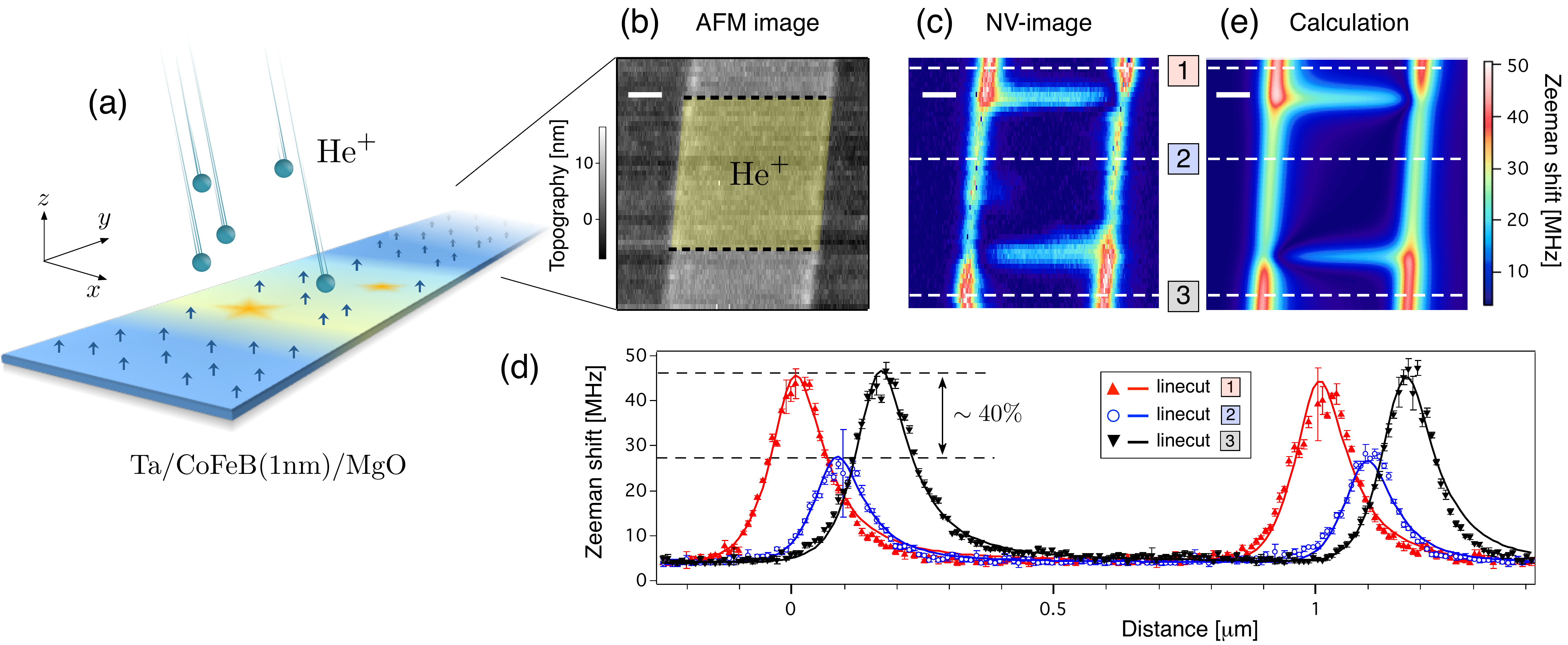}
\caption{(a) Schematic drawing illustrating the decrease in magnetic moment density induced by local irradiation with He$^+$ ions. (b) AFM image of a 1-$\mu$m-wide wire of Ta|CoFeB (1 nm)|MgO. The irradiation region is indicated in yellow color between the black dashed line. (c) Corresponding Zeeman-shift distribution recorded with scanning-NV magnetometry. The NV probe is characterized by $\theta=100^{\circ}$ and $\phi=48^{\circ}$. Scale bar : $200$~nm. (d) Linecuts extracted from the three white dashed lines in (c). The ratio between the field maxima above the edges directly indicates a relative decrease of $I_s$ by $\sim 40\%$ in the irradiated region. The solid lines are data fitting. (e) Calculation of the full Zeeman-shift distribution above the irradiated wire using the parameters extracted from the fit. Scale bar : $200$~nm.
}
\label{Fig4}
\end{center}
\end{figure*}

Local determination of $I_s$ through stray field mapping is not limited to magnetic samples with a 1D geometry, like a wire. The method can be easily extended to any type of nanostructured sample. This is illustrated by measuring the magnetic moment density of a 2D structure consisting of a $500$-nm-wide square dot etched in a Ta|CoFeB (1 nm)|MgO film [Fig.~\ref{Fig3}(a)]. The full Zeeman-shift distribution recorded above the square dot with a scanning-NV magnetometer is shown in Fig.~\ref{Fig3}(b). By fitting Zeeman-shift linecuts across the square, we infer once again $I_s = 98\pm 3 \ \mu_B/{\rm nm}^2$. Using this value, the full distribution was computed showing an excellent agreement with the experiment [Fig.~\ref{Fig3}(d)]. 

As a last experiment, we demonstrate how stray field imaging with scanning-NV magnetometry can be used as an efficient tool to analyze the uniformity of $I_s$ in ultrathin ferromagnets with submicron resolution. To this end, we investigate local modifications of $I_s$ induced by light irradiation with He$^+$ ions [Fig.~\ref{Fig4}(a)]. This method enables precise tuning of PMA and magnetization through intermixing at both Ta|CoFeB and CoFeB|MgO interfaces, and can be adjusted by varying the irradiation dose~\cite{Chappert1998,Devolder2013}. A 1-$\mu$m-wide wire of Ta|CoFeB (1~nm)|MgO was irradiated through a mask with He$^+$ ions at $15$~keV energy with a fluence of $1.6\times10^{15}$~ions/cm$^2$~\cite{SI}. An AFM image of the sample indicating the irradiated area is given in Fig.~\ref{Fig4}(b), while the corresponding Zeeman-shift distribution recorded with scanning-NV magnetometry is shown in Fig.~\ref{Fig4}(c). Two important features can be observed. First, a stray field is generated at the border of the irradiated window along the long axis ($y$) of the wire. This indicates an abrupt variation of $I_s$ induced by local irradiation. Second, the stray field at the edge of the wire is significantly lower in the irradiated region, corresponding to an overall decrease of $I_s$. To get more quantitative information, Zeeman-shift linecuts were extracted from the image [Fig.~\ref{Fig4}(d)]. By fitting the data, we infer a relative decrease of $I_s$ by $40\%$ in the irradiated area. We stress that for this particular experiment the exact knowledge of either $d$ or $I_s$ is not even required to infer the relative variation of $I_s$, since it is directly given by the ratio between the field maxima above the edge of the wire. The full Zeeman-shift distribution calculated with the parameters inferred from the fit reproduces very well all the characteristic features of the measurement~[Fig.~\ref{Fig4}(e)]. From this experiment, we can conclude that the irradiation process uniformly modifies the magnetic properties, at least on a length scale of $\approx 100$~nm corresponding to the spatial resolution of our measurement.

In conclusion, we have introduced a novel approach based on quantitative magnetic field imaging to infer locally and in a very sensitive fashion the magnetic moment density in ultrathin films with PMA. By employing a scanning-NV magnetometer, this method leads to absolute measurements of $I_s$ with a few percent uncertainty combined with a spatial resolution below $(100$~nm)$^2$. This corresponds to an improvement by more than four orders of magnitude compared to state-of-the-art techniques~\cite{Vernier2014}. The principle of the method being quite general, it could be extended to any kind of magnetization pattern by merely computing different equations used for data fitting. This method opens new perspectives for studying variations of magnetic properties at the nanoscale.
%Combining spatial resolution and accuracy with a non demanding environment, stray field measurement of $I_s$ could hence become a robust alternative to SQUID measurements to study ferromagnetic nano-structures of interest for spintronics, after their fabrication.\\

\indent The authors thank L. Rondin, J.-V. Kim, S. Rohart, A. Thiaville, T. Devolder, S. Eimer and L. Herrera Diez for experimental assistance and fruitful discussions. This research has been partially funded by the European Community's Seventh Framework Programme (FP7/2007-2013) under Grant Agreement n$^{\circ}$ 611143 (D{\sc iadems}) and n$^{\circ}$ 257707 (M{\sc agwire}), by C'Nano Ile-de-France through the project N{\sc anomag}, by the RTRA Triangle de la Physique and by the Labex NanoSaclay through the project S{\sc iltene}.

\begin{widetext}
\vspace{0.5cm}
\section{Supplementary Information} 

\subsection{Samples} \label{Sample}

The samples used in this work are stacks of Ta(5)$|$CoFeB(1)$|$MgO(2)$|$Ta(5) deposited on a Si$|$SiO$_2$(100~nm) substrate with a PVD Tamaris deposition tool by Singulus Tech (the number in brackets refers to the layer's thickness, expressed in nanometers). The stoichiometric composition of the as-deposited CoFeB layer is Co$_{20}$Fe$_{60}$B$_{20}$ for the data of Figs. 2 and 3, and Co$_{40}$Fe$_{40}$B$_{20}$ for the data of Fig. 4. The samples were patterned by e-beam lithography and ion milling to define magnetic wires (for Figs. 2 and 4) or dots (for Fig. 3). The etching depth $\delta d$, comprised typically between 10 and 50 nm, is larger than the depth of the ferromagnetic layer (7 nm), as illustrated in Fig. \ref{sample_cross-section}. Finally, a 100-nm-thick gold stripe was defined using a second step of lithography. This stripe is connected to a microwave generator and serves as an antenna to excite the NV center's spin resonances. Details about the scanning-NV magnetometry setup can be found in Ref.~[\onlinecite{Rondin2013}]

For the experiment described in Fig. 4, the sample was annealed at 300$^\circ$C for 2 hours, and a third step of e-beam lithography was used to open 1-$\mu$m-wide windows in a 400-nm-thick PMMA masking layer. The sample was then irradiated with helium ions, with an irradiation dose $F=1.6\cdot 10^{12}$ ions/cm$^2$ and an energy of 15.5~keV, after which the PMMA mask was removed.

\begin{figure}[t]
\begin{center}
\includegraphics[width=0.55\textwidth]{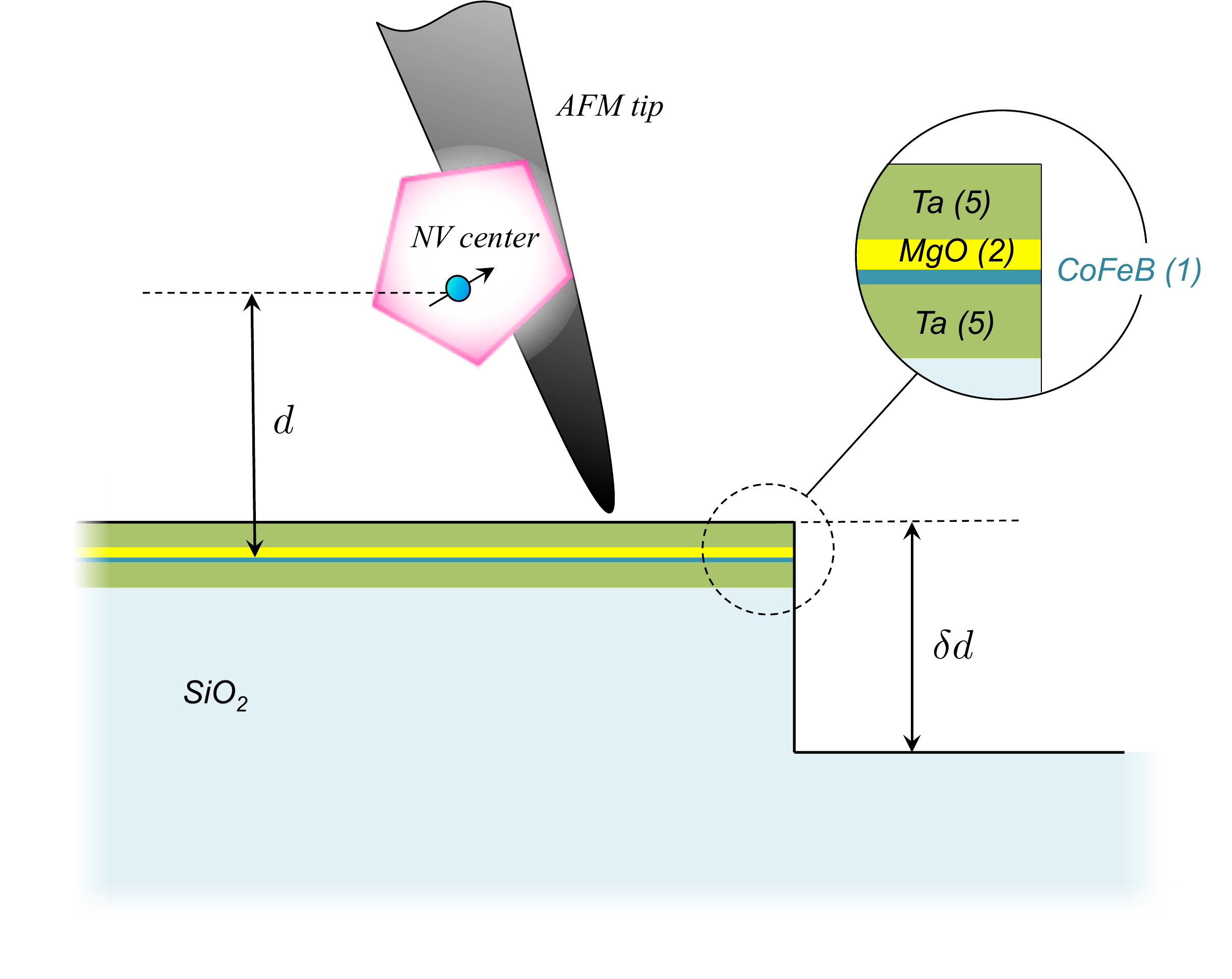}
\caption{
Schematic cross-section of the sample showing the layer stack and the etching depth $\delta d$. The distance between the NV center and the magnetic layer is denoted $d$.}
\label{sample_cross-section}
\end{center}
\end{figure}

\subsection{Derivation of the stray field above an edge of a perpendicularly magnetized film} \label{formula}

In this section, we derive the stray field above a semi-infinite layer of a ferromagnetic material with perpendicular magnetization. The layer lies in the $xy$ plane, has a thickness $t$ in the $z$ direction, and is bounded to the $x<0$ half space, with translation invariance along $y$. The saturation magnetization is denoted $M_s$.

As shown in Fig. \ref{Fig1}, such a magnetic layer may be seen as a capacitance, with magnetic charges on each surface. We can therefore make an analogy with electric field and electric charges. The charges are located on two half planes, one at $z=+\frac{t}{2}$ and one at $z=-\frac{t}{2}$. We thus start by computing a magnetic potential $V(P)$ created at the point $P$ by the charge distribution, such that the stray field $\vec{B}$ is defined as the gradient of the potential, $\vec{B}=-\vec{\nabla}V$. This potential is given by 
$$V(P) = \frac{\mu_0 M_s}{4\pi} \left\{ \int_{P'\in\mathit{S_{+\frac{t}{2}}}} \frac{dP'}{\left|\vec{r}_{P'P}\right|} - \int_{P'\in\mathit{S_{-\frac{t}{2}}}} \frac{dP'}{\left|\vec{r}_{P'P}\right|} \right\}~,$$ 
where $\left|\vec{r}_{P'P}\right|$ is the distance between the point $P$ at which we compute the field and a point $P'$ which belongs to the sample. In the frame of the magnetic layer, with $P$ of coordinates $(x,y,z)$ it reads 
$$V(x,\bcancel{y}, z) = \frac{\mu_0 M_s}{4\pi} \int_{-\infty}^{+\infty} dy' \int_{0}^{+\infty} dx' \left\{ \frac{1}{\sqrt{\left( x-x' \right)^2 + y'{} ^2 + \left( z-\frac{t}{2} \right)^2}} - \frac{1}{\sqrt{\left(x-x'\right)^2 + y'{}^2 + \left(z+\frac{t}{2}\right)^2}} \right\}~,$$ 
where there is no dependence in $y$ due to the translation symmetry.

\begin{figure}[t]
\begin{center}
\includegraphics[width=0.9\textwidth]{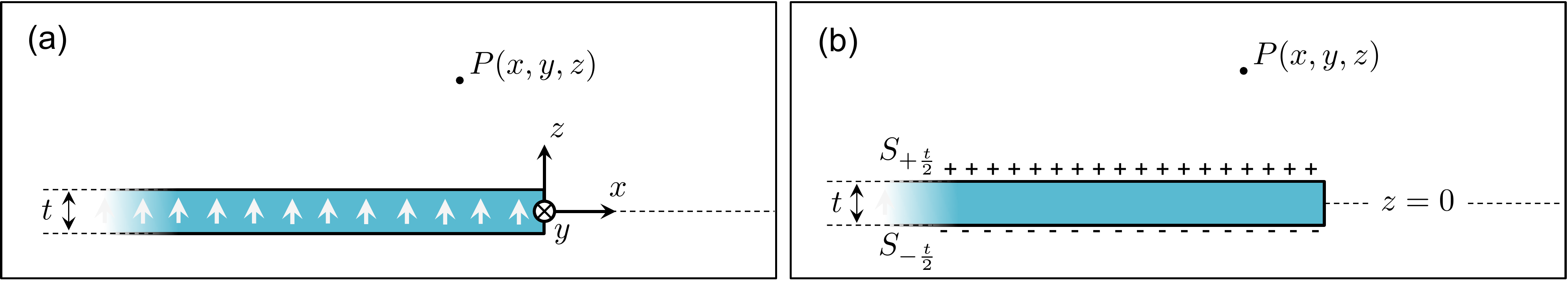}
\caption{(a) Geometry of the problem. A film of thickness $t$ with a uniform perpendicular magnetization presents an edge at $x=0$. The problem has translation invariance along the $y$ axis, and the origin is taken in the middle of the ferromagnetic layer. (b) Analogy with electrostatics. The perpendicularly magnetized film can be seen as two planes of magnetic charges, located at $z=\pm\frac{t}{2}$. From this charge distribution, one can compute a magnetic scalar potential $V(x,\bcancel{y},z)$, from which the magnetic field is retrieved.}
\label{Fig1}
\end{center}
\end{figure}

The integration over $x'$ gives
$$V(x,z)=\int_{-\infty}^{+\infty} dy' \left\{ \ln \left( \frac{\sqrt{x^2+y'^2+ \left( z-\frac{t}{2} \right) ^2+x}}{\sqrt{x^2+y'^2+ \left( z+\frac{t}{2} \right)^2+x}} \right) - \ln \left( \frac{y'^2+\left( z-\frac{t}{2} \right) ^2}{y'^2+ \left( z+\frac{t}{2} \right)^2} \right) \right\}~,$$ 
while the integration over $y'$ gives eventually
$$V(x,z)=\frac{\mu_0 M_s}{4\pi} \left\{  \left(z+\frac{t}{2}\right)\left[  \pi+2~{\rm atan}\left(  \frac{x}{z+\frac{t}{2}}  \right) \right] - \left(z-\frac{t}{2}\right)\left[  \pi+2~{\rm atan}\left(  \frac{x}{z-\frac{t}{2}}  \right) \right] \right. $$ $$ \left. + x\ln \left(  \frac{x^2+(z+\frac{t}{2})^2}{x^2+(z-\frac{t}{2})^2}  \right) \right\}~. $$

Finally, we obtain the stray field using the formula $\vec{B}=-\vec{\nabla}V$, which yields
\begin{equation}
\begin{dcases} 
B^{ed}_x(x,z) = \frac{\mu_0M_s}{4\pi}\ln\left(\frac{x^2+(z+\frac{t}{2})^2}{x^2+(z-\frac{t}{2})^2} \right) \\
B^{ed}_z(x,z) = \frac{\mu_0M_s}{2\pi} \left( {\rm atan}\left(\frac{x}{z+\frac{t}{2}}\right) - {\rm atan} \left(\frac{x}{z-\frac{t}{2}}\right) \right)
\end{dcases}~.
\end{equation} 

In the thin-film limit $z\gg t$, we can simplify these expressions into
\begin{equation}
\begin{dcases} 
B^{\rm ed}_x(x,z) = \ \ \frac{\mu_0M_s}{2\pi}~\frac{t}{z}~\frac{z^2}{x^2+z^2} + O((t/z)^3) \\
B^{\rm ed}_z(x,z) = -\frac{\mu_0M_s}{2\pi}~\frac{t}{z}~\frac{xz}{x^2+z^2} + O((t/z)^3)
\end{dcases}~,
\label{eq:B_edge} 
\end{equation} 
which is Eq. (1) of the main article when the product $M_st$ is replaced by $I_s$ and $z=d$.

\subsection{Fitting procedure} \label{fit}

In this section, we describe the fitting procedure used to retrieve the value of $I_s$ in a ferromagnetic wire.

When scanned above a perpendicularly magnetized wire, the NV center feels for each position $x$ a field $\vec{B}^w(x)$. This field is the contribution from the two edges of the wire and reads in the thin-film approximation
$$ \vec{B}^w(x) = - \vec{B}^{\rm ed}(x-x_L,z(x)) + \vec{B}^{\rm ed}(x-x_R,z(x))~, $$ 
where $\vec{B}^{\rm ed}(x,z)$ is defined by Eq. (\ref{eq:B_edge}), and $x_L$ ($x_R$) is the position of the left (right) edge of the magnetic wire. The height $z$ at which the field is probed by the NV depends on position $x$ according to the topography followed by the AFM tip, which resembles a top-hat function. The function $z(x)$ can be expressed as 
$$z(x)=d+{\rm topo}(x)~,$$ 
where $d$ is the average distance between the magnetic layer and the NV center when the AFM tip stands on top of the magnetic wire, and ${\rm topo}(x)$ corresponds to the measured topography, with an offset such that 
\begin{equation}
\begin{dcases} 
\overline{{\rm topo}(x)}=0~{\rm when~the~tip~stands~on~the~wire} \\
\overline{{\rm topo}(x)}=-\delta d~{\rm when~the~tip~stands~on~the~substrate}
\end{dcases}~,
\label{eq:topo} 
\end{equation} 
where $\overline{f(x)}$ designates the average over $x$ and $\delta d$ is the height of the etched wire. As illustrated in Fig. \ref{topo_shift}a, the functions $z(x)$ and ${\rm topo}(x)$ may be shifted in $x$ with respect to the actual sample's topography, hence with respect to the magnetic wire, because the NV center may be shifted with respect to the apex of the tip. This is the main reason for the difference in field amplitude associated with the two edges of the wire (see Fig. 2(b) of the main article), since the probe height $z(x)$ is in general different for the two edges at $x=x_L$ and $x=x_R$. A typical example of the ${\rm topo}(x)$ function extracted from AFM data is shown in Fig. \ref{topo_shift}b. 

\begin{figure}[t]
\begin{center}
\includegraphics[width=0.6\textwidth]{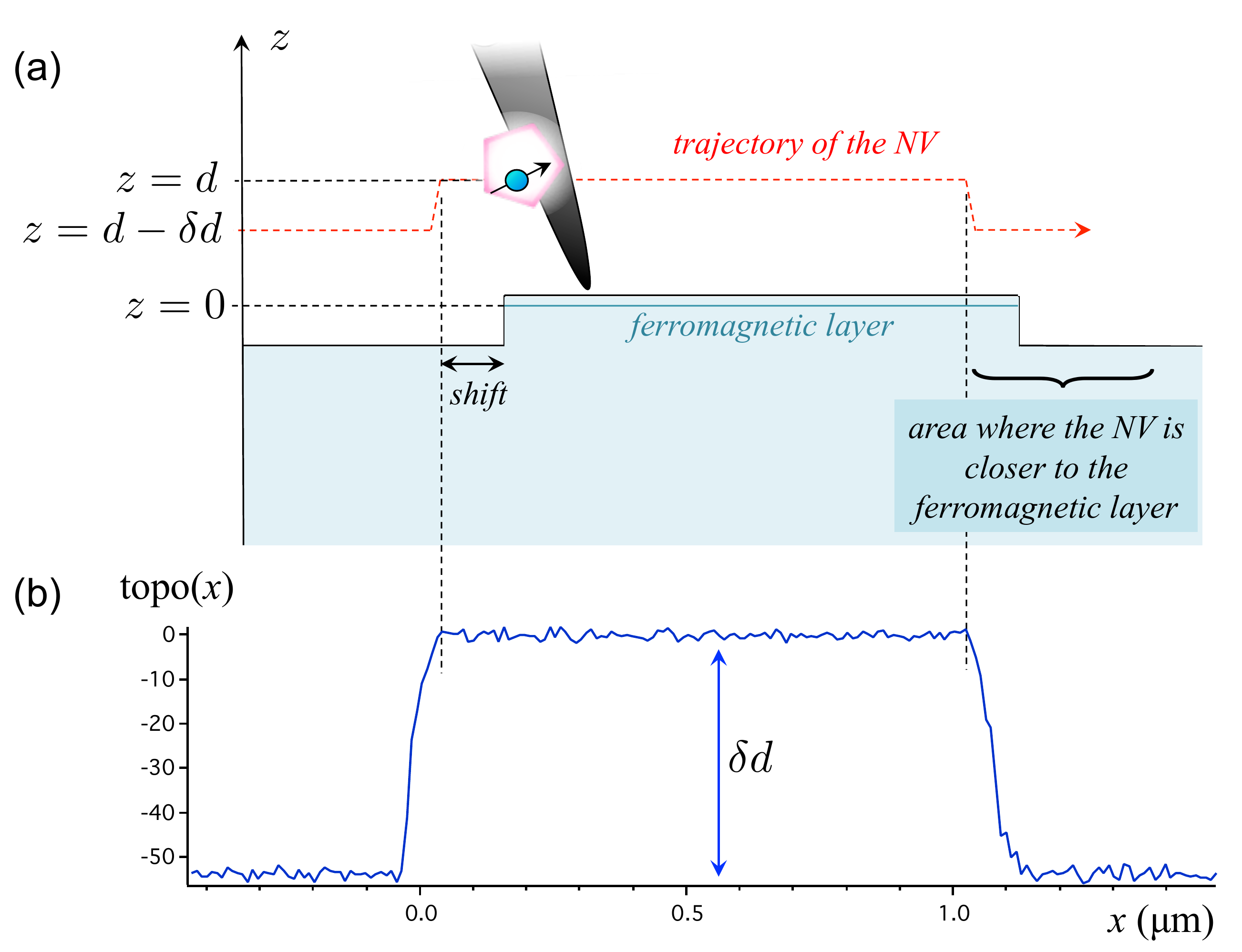}
\caption{(a) Scheme illustrating the shift between the measured AFM topography, hence the NV center's trajectory $z(x)=d+{\rm topo}(x)$, and the actual sample's topography, hence the position of the magnetic wire. (d) Example of ${\rm topo}(x)$ function measured on a Ta$|$Co$_{20}$Fe$_{60}$B$_{20}|$MgO wire. For this sample, the etching height is $\delta d\approx 53$ nm.}
\label{topo_shift}
\end{center}
\end{figure}

Note that since the scanners of our AFM are not equipped with position sensors due to volume constraints, and hence are somewhat imprecise, the topography data ${\rm topo}(x)$ together with the Zeeman shift data $\Delta f_{NV}(x)$ are first rescaled in order to match the width $w$ and height $\delta d$ of the magnetic wire as measured by a second, feedback-looped, calibrated AFM.

To obtain a fit function and compare it to experimental data, the magnetic field $\vec{B}^w(x)$ must be converted into a Zeeman shift $\Delta f_{NV}$ of the ESR frequencies of the NV center. Although it is well approximated in the present case by $\Delta f_{NV}=\sqrt{E^2 + (\gamma_e B_{NV}/2\pi)^2}$, where $B_{NV}=\vec{B}\cdot\vec{u}_{NV}$ is the projection of the magnetic field along the NV axis, we choose here to perform the exact computation by diagonalizing the Hamiltonian of the $S=1$ spin of the NV center
$$ \mathcal{H} = hDS_Z^2 + hE(S_X^2-S_Y^2) + h\frac{\gamma_e}{2\pi}\vec{B}\cdot\vec{S}~, $$
where $D$ and $E$ are the zero-field splitting parameters of the NV center, $h$ is the Planck constant, $\gamma_e/2\pi =$ 28.03(1)~GHz/T is the electron gyromagnetic ratio and $\vec{S}$ is the $S=1$ spin operator. The $XYZ$ frame is defined by the crystal orientation of the diamond, with $Z$ being along the axis $\vec{u}_{NV}$ of the defect, which is characterized by spherical angles $(\theta,\phi)$ in the $xyz$ lab frame~[\onlinecite{Rondin2013,Rondin2014}].

The resulting theoretical function $\Delta f_{NV}(x)$ is finally fitted to the data using least-squares minimization. The fit parameters are the stand-off distance $d$, the magnetic moment density $I_s$, and the positions $x_L$ and $x_R$ of the two wire's edges. The other parameters entering the fit function are the NV center's parameters ($D,E,\theta,\phi$) used in the Hamiltonian diagonalization, and the geometrical wire's parameters ($w,\delta d$) used for data rescaling. As mentioned in the main article and explained in details in Ref.~[\onlinecite{Tetienne2014bis}], the uncertainties for $d$ and $I_s$ are estimated based on the uncertainties of those six independently measured parameters ($D,E,\theta,\phi,w,\delta d$), as well as on the standard deviation among a series of measurements.

\section{Summary of the results} \label{results}

The magnetic moment density $I_s$ found for the various samples studied in this work are gathered in Table \ref{Table2}. Note that the value found for the sample irradiated through the PMMA masking layer [Fig. 4 of the main paper] is significantly smaller than the one reported in Ref.~[\onlinecite{Vernier2014}] for non-irradiated samples. This is due to the fact that a 400-nm PMMA layer does not completely block 15.5 keV helium ions, as confirmed by SRIM simulation. \\

\begin{table}[h]
\begin{center}
\begin{tabular}{|c|c|c|c|c|}
\hline 
\hline 
~~Sample~~ & ~~~~~Co$_{20}$Fe$_{60}$B$_{20}$~~~~~ & ~~~~~Co$_{40}$Fe$_{40}$B$_{20}$~~~~~ & ~~~~~Co$_{40}$Fe$_{40}$B$_{20}$~~~~~ & ~~~~~Co$_{40}$Fe$_{40}$B$_{20}$~~~~~ \\
& as deposited & as deposited & annealed and & annealed and \\
& & & irradiated & irradiated through \\
& & & & 400-nm PMMA \\
\hline
~~~$I_s$ ($\mu_B$/nm$^2$)~~~ & $97.7 \pm 3.0 $ & $97.8 \pm 1.9 $ & $42.5 \pm 2.0$  &  $72.9 \pm 2.0 $  \\
\hline
\hline 
\end{tabular}
\caption{
Value of $I_s$ measured with NV magnetometry on different magnetic wires.
}
\label{Table2}
\end{center}
\end{table}

\end{widetext}


\begin{thebibliography}{99}
\bibitem{Note}
In this work ,we use the Bohr magneton as magnetic unit, $\mu_B=9.27\times 10^{-24}$~A.m$^2$.
\end{thebibliography}
\end{document}